\documentclass[%
 reprint,
showpacs,preprintnumbers,showkeys,
 amsmath,amssymb,
 aps,
]{revtex4-1}

\usepackage{graphicx}
\usepackage{dcolumn}
\usepackage{bm}
\usepackage[caption=false]{subfig}

\usepackage{fancyhdr}
\usepackage[colorlinks,linkcolor=blue]{hyperref}

\begin{document}


\title{Graphene based functional devices: A short review}

\author{Rong Wang}
\affiliation{Department of Electrical and Electronic Engineering, The University of Hong Kong, Pokfulam Road, Hong Kong.}
\author{Xingang Ren}
\email{xgren@ahu.edu.cn}
\affiliation{Key Laboratory of Intelligent Computing \& Signal Processing, Ministry of Education, Anhui University,  Hefei 230039, China}
\affiliation{Department of Electrical and Electronic Engineering, The University of Hong Kong, Pokfulam Road, Hong Kong.}
\author{Ze Yan}
\affiliation{School of Instrumentation Science \& Opto-electronics Engineering, Beihang University, Beijing 100191, China.}
\author{Lijun Jiang}
\email{ljiang@eee.hku.hk}
\affiliation{Department of Electrical and Electronic Engineering, The University of Hong Kong, Pokfulam Road, Hong Kong.}
\author{Wei E. I. Sha}
\affiliation{Key Laboratory of Micro-nano Electronic Devices and Smart Systems of Zhejiang Province, College of Information Science \& Electronic Engineering, Zhejiang University, Hangzhou 310027, China.}
\author{Guangcun Shan}
\email{gcshan@buaa.edu.cn}
\affiliation{School of Instrumentation Science \& Opto-electronics Engineering, Beihang University, Beijing 100191, China.}
\affiliation{California NanoSystem Institute and Department of Chemistry \& Biochemistry, University of California, Los Angeles, CA 90095,USA}
%
%

\date{\today}

\begin{abstract}
Graphene is an ideal 2D material system bridging electronic and photonic devices. It also breaks the fundamental speed and size limits by electronics and photonics, respectively. Graphene offers multiple functions of signal transmission, emission, modulation, and detection in a broad band, high speed, compact size, and low loss. Here, we will have a brief view of graphene based functional devices at microwave, terahertz, and optical frequencies. Their fundamental physics and computational models will be discussed as well.
\end{abstract}



\maketitle

\section{Introduction}
Electron wavelength is much shorter than photon wavelength. Hence electronic devices exhibit compact sizes. Unfortunately, the electron transport is limited by its mobility or scattering events, which lowers its speed compared to photon transmission or propagation. On the other hand, the sizes of photonic devices are held back by the diffraction or half-wavelength limit. Surface plasmon polariton, which is an entangled state of electron and photon, is a promising candidate to achieve high speed and small size simultaneously. Unfortunately, high ohmic loss of metallic plasmon devices creates a fundamental hurdle for their potential applications.

Graphene, a conjugated carbon sheet arranged in a two-dimensional (2D) hexagonal lattice, allows multifunctions in signal emission, transmission, modulation, and detection, featured with the broad band, high speed, compact size, and particularly low loss. Compared to traditional materials such as silicon and III-V semiconductors, graphene demonstrates unique properties in its high electron mobility, large thermal conductivity, strong mechanical ductility, and high third-order optical nonlinearities. There are many reviews in literature about graphene photonics and plasmonics \cite{Neto,Geim,Allen,Avouris,Bonaccorso,Bao,Grigorenko,Novoselov,Tassin,Abajo1,Abajo2,Low,Otsuji,Stauber,Emani,Xiao,Huang,Gusynin}. Here, we will present a brief review of graphene based functional devices at microwave, terahertz, and optical frequencies. The paper is organized as follows. The basic physical foundations and computational models of graphene will be discussed in Section 2 first. Then, the fabrication and relevant experimental results of graphene based functional devices at microwave frequencies will be introduced in Section 3. In Section 4, the application of graphene in dynamically altering the antenna characteristics is presented from terahertz to mid-infrared bands. Section 5 will present recent progresses on graphene modulators at terahertz and infrared bands. Then graphene detectors at terahertz frequencies will be discussed in details at Section 6. At the end, Section 7 will provide a brief summary for graphene based functional devices.

\section{Physical foundation and theoretical modeling}
\begin{figure}[h]
\includegraphics[width = 0.3\textwidth]{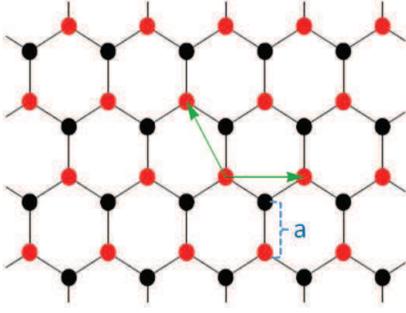}
\caption{\label{fig:1} Hexagonal lattice of graphene. Two atoms or sub-lattices per unit cell are denoted by red and black, respectively. The green arrows show two lattice vectors. The length of the nearest-neighbor bond is $a$.}
\end{figure}

Graphene is a hexagonal lattice of carbon atoms arranged in a single atomic sheet. The electronic structure of graphene is essential to its electromagnetic responses. By using tight-binding model and creation and annihilation operators, the Hamiltonian of graphene can be written as
\begin{equation}\label{eq1}
H =  - t\sum\limits_i {a_{{{\bf{r}}_i}}^ + b_{{{\bf{r}}_i} + {{\bf{e}}_1}}^{}}  - t\sum\limits_i {a_{{{\bf{r}}_i}}^ + b_{{{\bf{r}}_i} + {{\bf{e}}_2}}^{}}  - t\sum\limits_i {a_{{{\bf{r}}_i}}^ + b_{{{\bf{r}}_i} + {{\bf{e}}_3}}^{}}  + h.c.
\end{equation}
where ${{\bf{e}}_1} = (0,a),{\rm{ }}{{\bf{e}}_2} = \left( {\frac{{ - \sqrt 3 a}}{2},\frac{{ - a}}{2}} \right),{\rm{ }}{{\bf{e}}_3} = \left( {\frac{{\sqrt 3 a}}{2},\frac{{ - a}}{2}} \right)$ are the three vectors pointing to three nearest-neighbor sites. $a$ is the length of the nearest-neighbor bonds. $t\approx2.8$ eV is the nearest-neighbor hopping energy. The above Hamiltonian can be recasted into a series of summations in the momentum space, i.e.
\begin{equation}\label{eq2}
\begin{array}{l}
H =  - t\sum\limits_{\bf{k}} {a_{\bf{k}}^ + b_{\bf{k}}^{}} ({e^{ - i{\bf{k}} \cdot {{\bf{e}}_1}}} + {e^{ - i{\bf{k}} \cdot {{\bf{e}}_2}}} + {e^{ - i{\bf{k}} \cdot {{\bf{e}}_3}}}) + h.c.\\
 = \sum\limits_{\bf{k}} {\left( {\begin{array}{*{20}{c}}
{a_{\bf{k}}^ + }&{b_{\bf{k}}^ + }
\end{array}} \right)} \left( {\begin{array}{*{20}{c}}
0&{{H_{12}}({\bf{k}})}\\
{{H_{21}}({\bf{k}})}&0
\end{array}} \right)\left( {\begin{array}{*{20}{c}}
{a_{\bf{k}}^{}}\\
{b_{\bf{k}}^{}}
\end{array}} \right)
\end{array}
\end{equation}
where ${H_{12}}({\bf{k}}) =  - t({e^{ - i{\bf{k}} \cdot {{\bf{e}}_1}}} + {e^{ - i{\bf{k}} \cdot {{\bf{e}}_2}}} + {e^{ - i{\bf{k}} \cdot {{\bf{e}}_3}}}) = H_{21}^ * ({\bf{k}})$. The eigenvalues of Hamiltonian lead to the dispersion relation
\begin{equation}\label{eq3}
\begin{split}
&E({\bf{k}}) =  \pm \sqrt {H_{12}^{}({\bf{k}})H_{12}^*({\bf{k}})}  \\
&=  \pm t\sqrt {1 + 4\cos \left( {\frac{{\sqrt 3 a}}{2}{k_x}} \right)\cos \left( {\frac{{3a}}{2}{k_y}} \right) + 4{{\cos }^2}\left( {\frac{{\sqrt 3 a}}{2}{k_x}} \right)}
\end{split}
\end{equation}
The conduction and valance bands touch each other at the six Dirac points of the first Brillouin zone. Using Taylor expansion at the Dirac points, the dispersion relation of graphene is approximate to be linear, i.e.
\begin{equation}\label{eq4}
E({\bf{k}}) =  \pm \hbar {v_F}\left| {\bf{k}} \right|,\quad \hbar {v_F}{\rm{ = 3}}at/2
\end{equation}
where ${v_F} \approx {10^6}$ m/s is the Fermi velocity.

The electron motion of graphene at the conduction band, which is also known as the intraband transition, can be described by the Boltzmann equation
\begin{equation}\label{eq5}
\frac{{\partial f}}{{\partial t}} + \frac{{\partial {\bf{p}}}}{{\partial t}} \cdot \frac{{\partial E}}{{\partial {\bf{p}}}}\frac{{\partial f}}{{\partial E}} = \frac{{{f_0} - f}}{\tau }
\end{equation}
where $f$ is a probability density function of electron defined in the $(\mathbf{r},\mathbf{k})$-phase space. $E$ and $\mathbf{p}$ are the kinetic energy and momentum of electron. Here we use the relaxation time approximation and $\tau$ is the relaxation time. Moreover, we ignore the term $(\partial \mathbf{r}/\partial t)(\partial f/\partial \mathbf{r})$ in the Boltzmann equation, which is related to the diffusion current ignored for the conduction electrons. The above formulation can be rewritten as
\begin{equation}\label{eq6}
\frac{{\partial f}}{{\partial t}} + {\bf{F}} \cdot {\bf{v}}\frac{{\partial f}}{{\partial E}} = \frac{{{f_0} - f}}{\tau }
\end{equation}
where $\bf{v}$ is the group velocity and $\mathbf{F}=e\mathbf{E}$, where $\mathbf{E}$ is the electric field. Moreover, $E=\hbar \omega,\,{\rm{ }}{\bf{p}}=\hbar {\bf{k}},\,{\rm{ }}{\bf{v}}= \partial \omega /\partial {\bf{k}}$. Using the Fourier transform, we arrive at
\begin{equation}\label{eq7}
\left( { - i\omega  + \frac{1}{\tau }} \right)f =  - {\bf{F}} \cdot {\bf{v}}\frac{{\partial f}}{{\partial E}} + \frac{{{f_0}}}{\tau }
\end{equation}
By multiplying Eq. \eqref{eq7} by $\bf{v}$ and then integrating it over $\bf{k}$ space, and we finally get
\begin{equation}\label{eq8}
\begin{split}
&\frac{g}{{{{\left( {2\pi } \right)}^2}}}\left( { - i\omega  + \frac{1}{\tau }} \right)\int {ef{\bf{v}}d{k_x}d{k_y}}  =\\
&\frac{{eg}}{{{{\left( {2\pi } \right)}^2}}}\int { - {\bf{F}} \cdot \left( {{\bf{vv}}} \right)\frac{{\partial f}}{{\partial E}}} d{k_x}d{k_y}
\end{split}
\end{equation}
where $g$ is degeneracy, which is equal to $4$ for the graphene sheet considering spin degeneracy and two carbon atoms in a unit cell.

By using ${\bf{J}} = {g/{{(2\pi )}^2}} \int {ef{\bf{v}}d{k_x}d{k_y}} $, we can get the expression for the conduction-band current
\begin{equation}\label{eq9}
{\bf{J}} = \frac{1}{{ - i\omega  + \frac{1}{\tau }}}\frac{e}{{{\pi ^2}}}\int {{\bf{F}} \cdot \left( {{\bf{vv}}} \right)\frac{{ - \partial f}}{{\partial E}}} d{k_x}d{k_y}
\end{equation}
Next, we consider the electrical impedance and ignore the magnetic field that is responsible for the Hall impedance, i.e.
\begin{equation}\label{eq10}
\sigma _{xx}^{{\mathop{\rm int}} ra} = \frac{1}{{ - i\omega  + \frac{1}{\tau }}}\frac{{{e^2}}}{{{\pi ^2}}}\int {v_x^2\frac{{ - \partial f}}{{\partial E}}} d{k_x}d{k_y} = \sigma _{yy}^{{\mathop{\rm int}} ra}
\end{equation}
By using the approximate linear dispersion relation Eq. \eqref{eq4} and replacing $v_x^2$ with $v_F^2/2$, in view of the fact $\partial f/\partial E$ is small except in the vicinity of Fermi-level. As a result, Eq. \eqref{eq10} can be rewritten as
\begin{equation}\label{eq11}
\begin{split}
\sigma _{xx}^{{\mathop{\rm int}} ra} &= \frac{1}{{ - i\omega  + \frac{1}{\tau }}}\frac{{{e^2}}}{{\pi {\hbar ^2}}}\int {\frac{{ - \partial f}}{{\partial E}}E} dE\\
  &=\frac{{{\rm{ - }}i}}{{\omega  + i\frac{1}{\tau }}}\frac{{{e^2}}}{{\pi {\hbar ^2}}}\int_0^\infty  E \left( {\frac{{\partial f(E)}}{{\partial E}} - \frac{{\partial f( - E)}}{{\partial E}}} \right)dE
\end{split}
\end{equation}
where $f = 1/({e^{(E - {\mu _c})/{k_B}T}} + 1)$ is the Fermi-Dirac distribution, and $\mu _c$ is the chemical potential. The above agrees with the intraband transition part of Kubo formula \cite{Gusynin}. The Eq. \eqref{eq11} is responsible for the imaginary part of the complex conductivity and thus plasmonic effects.

The real part of the conductivity, which corresponds to the interband transition and generation of electron-hole pair, can be obtained by employing the kramers-kronig relation or Fermi's golden rule \cite{Jablan}. Finally, the complex surface conductivity is of the following form \cite{Gusynin}
\begin{equation}\label{eq12}
\begin{split}
{\sigma ^s} = \frac{{{\rm{ - }}i}}{{\omega  + i\frac{1}{\tau }}}\frac{{{e^2}}}{{\pi {\hbar ^2}}}\int_0^\infty  E \left( {\frac{{\partial f(E)}}{{\partial E}} - \frac{{\partial f( - E)}}{{\partial E}}} \right)dE{\rm{ + }}\\
i\left( {\omega  + i\frac{1}{\tau }} \right)\frac{{{e^2}}}{{\pi {\hbar ^2}}}\int_0^\infty  {\frac{{f( - E) - f(E)}}{{{{\left( {\omega  + i/\tau } \right)}^2} - 4{{\left( {E/\hbar } \right)}^2}}}} dE
\end{split}
\end{equation}
When photon energy is larger than $2\mu_c$, optical conductivity is dominated by the interband transition. As carrier concentration increases by the electrostatic doping of graphene such that the photon energy is less than $2\mu_c$, intraband transition becomes dominant while interband transitions are suppressed due to the Pauli blocking.

There are two common scenarios to theoretically model graphene by classical computational electromagnetics algorithms. One approach is to use the effective bulk permittivity. The in-plane bulk permittivity of graphene is defined as ${\varepsilon _r} = 1 + i{\sigma ^s}/(\omega {\varepsilon _0}d)$, where $d$ refers to the effective thickness of graphene sheet, and $\sigma ^s$ is the surface conductivity calculated by the Kubo-formula of Eq. \eqref{eq12}. The out-of-plane permittivity is chosen to be the dielectric permittivity of $2.5$ \cite{Yao}. Another approach is to adopt the surface conductivity directly; and implement corresponding surface impedance condition \cite{Nayyeri,Li1,Li2}, i.e. ${\bf{n}} \times \left( {{{\bf{H}}_1} - {{\bf{H}}_2}} \right) = {\sigma ^s}{{\bf{E}}_t}$, where ${{\bf{H}}_1}$ and ${{\bf{H}}_2}$ are the magnetic fields above and below the graphene sheet and ${{\bf{E}}_t}$ is the electric field tangential to the graphene sheet. More details regarding computational techniques can be found in \cite{Shao}.

\section{Fabrication and microwave devices}
Tremendous efforts have been made over the last decade towards the fabrication of graphene, which, in general, can be categorized into three groups: (1) mechanical cleavage (MC), (2) epitaxial growth (EG), and (3) chemical vapor deposition (CVD). MC is realized by mechanically splitting bulk graphite into atomically thin single layer graphene with the aid of an adhesive tape. The first in-lab fabrication of single layer graphene was demonstrated by Novoselov et al. in 2005 \cite{Novoselov} and the ballistic mobilities of up to $10^{6}\,\,\mathrm{cm}^{2}\,\mathrm{V}^{-1}\,\mathrm{s}^{-1}$ were experimentally observed \cite{mobility1, mobility2}. Although MC is favorable for high-quality graphene fabrication, the very limited size (often coexists with other multiple atomic layers) and the time-cost exfoliation process prevent the large-scale industrial production using this method. Therefore, the artificial synthesis of scaled-up single layer graphene has received great interests all over the world. EG, also known as growth on silicon carbide, utilizes high temperature (over 1000 $^\circ$C) to thermally evaporate silicon (0001) from silicon carbide and leaves graphene films on the surface \cite{SiC1, SiC2}. The mobility ranging from  $1\times 10^{4}\,\,\mathrm{cm}^{2}\,\mathrm{V}^{-1}\,\mathrm{s}^{-1}$ to $3\times 10^{4}\,\,\mathrm{cm}^{2}\, \mathrm{V}^{-1}\,\mathrm{s}^{-1}$ was reported on the EG-grown graphene surface measured at room temperature \cite{SiC3,SiC4}. Since silicon carbide has been widely used in high-speed electronics as a bottom substrate, the direct growth of single layer graphene on silicon carbide circumvents the additional substrate transfer process and benefits the graphene-integrated silicon carbide devices such as high-frequency transistors \cite{transistor} and light emitting diodes \cite{LED}. However, due to the non self-limiting nature of thermal decomposition, it's difficult to generate pure single-layer carbon surface on silicon carbide surface and special care needs to be paid during the EG synthesis process. Furthermore, the high cost of silicon carbide wafers (more than 100 US dollar for a 4-inch wafer) limits the massive manufacture of graphene. In order to fulfil the low-cost and large-scale single layer graphene fabrication, CVD steps onto the stage, which, in principle, decomposes carbon atoms from hydrocarbons (methane, for example) and nucleates them on certain kinds of metal surfaces such as copper and finally forms an uniform and large area single layer graphene. The first large-area synthesis of single layer graphene using CVD was demonstrated in 2009 \cite{CVD} and mobility at room temperature varying from $1.64\times 10^{4}\,\,\mathrm{cm}^{2}\,\mathrm{V}^{-1}\,\mathrm{s}^{-1}$ to $2.5\times10^{4}\,\,\mathrm{cm}^{2}\,\mathrm{V}^{-1}\,\mathrm{s}^{-1}$ was obtained after the transfer of the single-layer graphene from the copper foil to the insulated substrate \cite{CVD1,CVD2}. Different from EG, CVD is almost self-contained. It means that the growth procedure will automatically pause when the entire catalytic metal surface is covered with single layer graphene. As a result, large area (over square decimeters) and uniform single layer graphene can be efficiently synthesized using this method. However, by virtue of the difference in thermal expansion coefficient between copper and graphene \cite{CVD}, defective wrinkles could occur on the graphene surface. They significantly degrade this two-dimensional material's quality, which can be seen from the much smaller mobility compared with the ones of MC-graphene. In order to provide cost-effective as well as high-quality single layer graphene for massive production, further optimization and improvement need to be done for the CVD method.

\begin{figure}[!b]
\centerline{\includegraphics[scale=0.48]{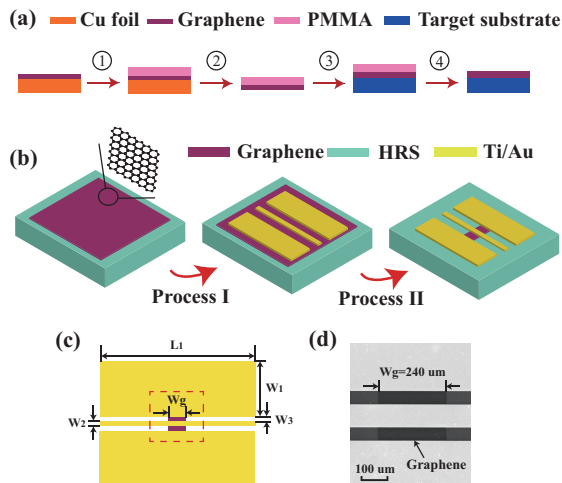}}
\caption{(a) Graphene transfer flow charts. (b) Fabrication process flow for metal deposition and graphene patterning. (c) Layout of the graphene-integrated coplanar waveguide. (d) Scanning electron microscope (SEM) images for the dashed red-line region in (c), where the graphene flakes of $48\times240\,\,\mathrm{\mu m}^2$ are placed.}
\label{transfer}
\end{figure}

Following the growth of graphene on copper foil in CVD chamber, the single layer graphene shall be transferred to the desired substrates with the aid of PMMA (Polymethyl Methacrylate) \cite{PMMA}. As illustrated in Fig. \ref{transfer}(a), this transfer process is composed of four steps: (1) A PMMA layer is firstly spin coated on top of graphene layer; (2) The supporting copper is etched away using Cu etchant; (3) The PMMA/graphene membrane is then attached to the target substrate such as the high-resistivity silicon; (4) The PMMA adhesive layer is dissolved in organic solvent and the graphene is eventually left on the target substrate. A distilled water rinse is suggested to get rid of the chemical residues during the transfer process. The adhesiveness between graphene and substrate could be further enhanced with several hour's baking so that the water molecules can fully evaporate.

A graphene-integrated circuit is ready for manufacture after the successful transfer process. Here, we take the graphene-embedded coplanar waveguide as an example and introduce the fabrication technologies utilized for electrodes deposition and graphene patterning \cite{process_flow}. As shown in Fig. \ref{transfer}(b), the deposition of coplanar waveguide electrodes (Process I) includes three stages: (1) Photo-lithography is applied to form the coplanar waveguide pattern on the photoresist layer; (2) Bi-metal layer (20 nm Ti/100 nm Au) is then deposited on the graphene sheet using E-Beam evaporation; (3) The unwanted metal is removed during lift-off. Graphene patterning (Process II) is also achieved by three steps: (1)
A pair of $48\times240\,\,\mathrm{\mu m}^2$ Photoresist pattern cover on the graphene sheet between the signal line and the ground; (2) the graphene sheet that is not covered by photoresist is etched away by the oxygen plasma. Note that the plasma power cannot be too high to destroy the protection photoresist; (3) Photoresist is finally stripped with the acetone followed by distilled water rinse. Fig. \ref{transfer}(c) illustrates the layout of the graphene-embedded coplanar waveguide and the patterned graphene flakes can be clearly identified under scanning electron microscope (SEM) in Fig. \ref{transfer}(d).

The development of CVD enables the fabrication of large-area graphene with the size comparable to the wavelength of microwave frequencies and boosts the usage of graphene at microwave band. Before this, most of the works were focused on the simulation region, such as graphene antenna \cite{antenna}, orbital angular momentum modulator \cite{OAM} and frequency selective surface \cite{FSS}. The simulated graphene layer is normally treated as an impedance surface or an ultra-thin bulk layer based on Kubo formula \cite{kubo}. Partial element electrical circuit, a derived circuit-level modeling method, has also been utilized for simulations \cite{PEEC}. However, as mentioned in the previous section, the quality of the CVD-graphene cannot reach the ideal expectation at this stage. It hinders the realization of many theoretical simulations which employ graphene as a conductive surface. Several works, on the other hand, take the advantage of the lossy characteristic of CVD-graphene and propose the graphene-embedded microstrip attenuator \cite{attenuator} and tunable transmission line from experimental results \cite{tunable}.

\section{Tunable graphene-based antennas from terahertz to mid-infrared}
The monolayer graphene allows to be p/n-doped under positive/negative electrostatic bias. The injection of charge carriers will shift the graphene chemical potential (Fermi level) away from the Dirac point resulting in the tunable surface conductivity. Due to the forbidden interband transitions by the Pauli blocking, the graphene monolayer reveals the metallic behavior and supports strong plasmonic effects from the terahertz to mid-infrared even to visible light regions. More importantly, the graphene monolayer with tunable chemical potential (through modulating the gated voltage) enables us to gradually engineer its surface conductivity and subsequently manipulate the graphene plasmon for realizing the tunable optical response of graphene-based antennas. In this section, we will discuss the graphene monolayer as a key element for dynamically altering the antenna characteristics.

\subsection{Manipulation of graphene absorption}
Regarding the reciprocity theorem, the graphene-based antennas are capable of receiving and transmitting electromagnetic waves. As the receiver, besides the radiation pattern, gain and impedance, the optical absorption plays an important role in converting the propagating wave into surface wave toward high responsivity and also in enhancing near-field. The modulation of the absorption of graphene-based antenna will offer in-depth insight on how to engineer the antenna characteristics.

As we know, the intrinsic value of optical absorption of undoped monolayer graphene is 2.3\%, which is insufficient to realize strong light-matter interaction. The graphene plasmon and photonic modes, critical coupling effects, etc, have been utilized to improve the absorption of graphene antennas. For instance, through modulating the graphene chemical potential, the log-periodic toothed antenna made by graphene has delivered the multi-resonances with highly tunable intensity and spectral location \cite{RenRef1}. The active tunable absorption has been demonstrated in the closely packed graphene nanodisk arrays, in which the disk geometrical parameters, interparticle spacing and voltage-driven electrostatic doping can be optimized to achieve 30\% graphene absorption with one order of magnitude enhancement as compared to that of the planar un-doped graphene monolayer \cite{RenRef2}. In a subsequent work, Halas et. al. also demonstrated the plasmon energy and strength can be tuned in the nanostructured graphene monolayer disk array \cite{RenRevRef3}. Theoretically, the critical coupling effect has been realized to attain the perfect absorption of graphene, the combined guided mode of supstrate dielectric grating and photonic bandgap of substrate Bragg grating cooperatively contribute to near-unity absorption of the graphene \cite{RenRef3}.

Besides, the integrations of graphene monolayer with various types of metallic structures also offer additional degrees of freedom for engineering the graphene absorption. Qin et al. desined the antenna designs to enhance the interaction between metallic structures and graphene layer. The graphene monolayer with 30\% absorption has been achieved in an ultra-broadband spectral range from 780 nm to 1760 nm through integrating two types of the split cross antennas \cite{RenRef4}. In addition to the absorption, the experimental results showed that the blackbody emission of graphene-based resonator can also be electrically modulated \cite{RenRef5}.

\subsection{Modulation of optical characteristics}
\subsubsection{utilization of tunable graphene conductivity}
The electrostatic bias on graphene can gradually change its dielectric constant, which plays an important role in continuously modulating the antenna characteristics \cite{RenRef6, RenRef7,RenRef8}. The remarkable modulation depth up to 90\% of transmission has been demonstrated in the graphene-loaded silver ribbon antenna via turn on- and off-resonance \cite{RenRef9}. Yu and coworkers showed the large modulation of both the amplitude and phase in graphene-metal antenna. The intensity modulation ratio of 100 and phase modulation of $240^\circ$ have been demonstrated through dynamically tuning the graphene surface conductivity \cite{RenRef10}. The control of the magnetic resonance of diabolo antenna by integrating graphene monolayer has been shown with the resonance tuning range up to 63\% and intensity modulation up to 1460\% in mid-infrared wavelength range \cite{RenRef11}. Capasso and coworkers demonstrated the \textit{in situ} control of graphene-loaded mid-infrared antenna through electrically tuning the applied gated voltage, the electrostatically gated graphene located at the antenna gap enables the change of wavelength range up to 600 nm with the modulation depth of the intensity more than 30\% \cite{Yao}. They further demonstrated a large tuning range of 1100 nm (80\% of bandwidth) at the mid-infrared region by incorporation of metal-insulator-metal waveguide design in the graphene-based antenna \cite{RenRef13}. In another work, they studied the electrically tunable absorber composed of Fabry-Perot antenna and graphene. The gated voltage applied on the graphene enables to modulate in and out of the critical coupling condition, the modulation depth up to 100\% and speed of 20 GHz have been realized over a broadband wavelength range from the near-infrared to terahertz wavelength \cite{RenRef14}.

\begin{figure*}[!htbp] 
	\centerline{
		\includegraphics[width=6.6in]{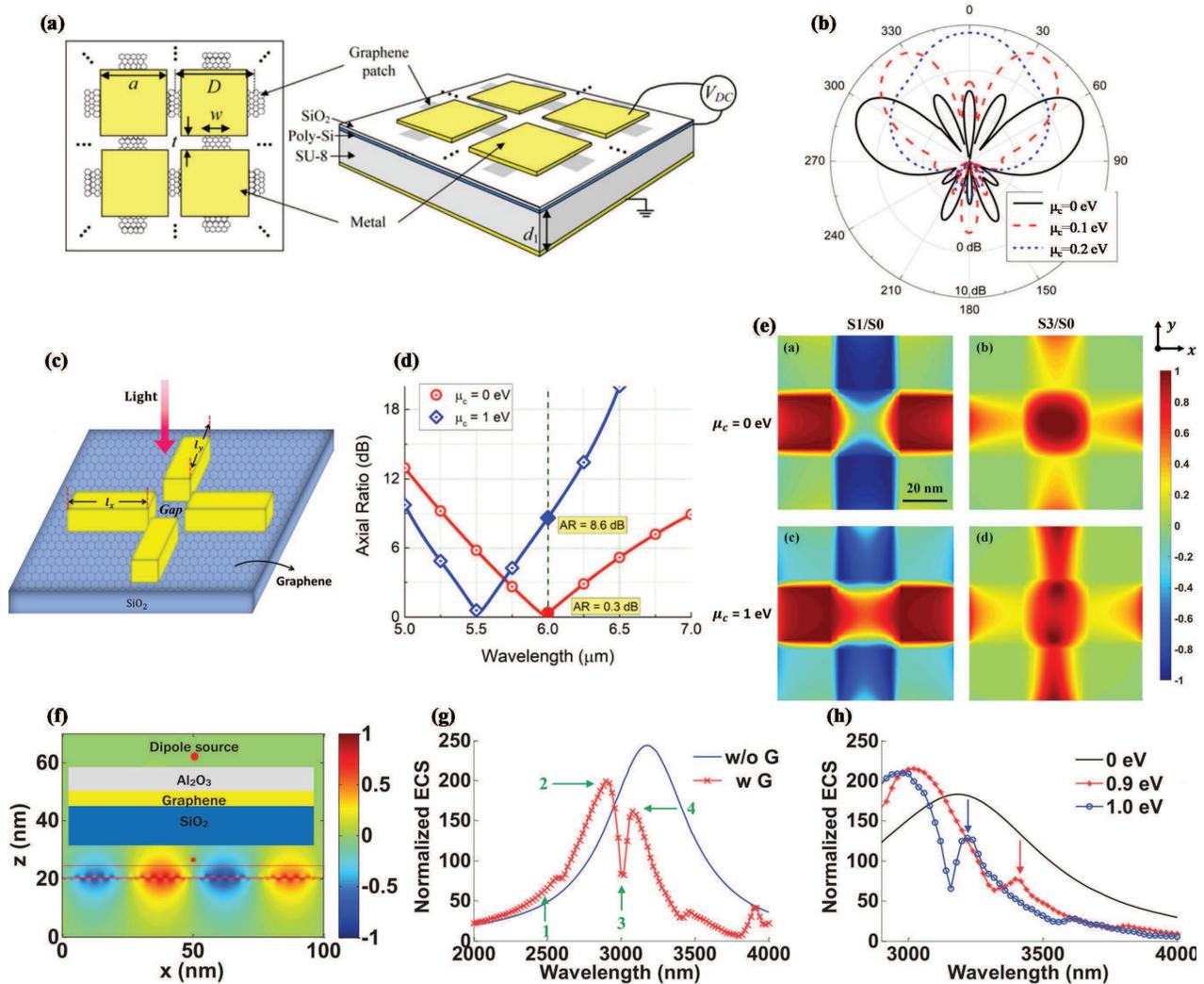}
	}
	\caption{ (Color online)  (a) Schematic of the 2D leaky-wave antenna with the graphene as a high-impedance surface, (b) E-plane radiation patterns with different chemical potential $\mu_c$ \cite{RenRef15}. (c) Schematic of the unit cell of the proposed graphene-loaded electrically tunable polarizer, (d) axial ratio of the reflected beam of the cross antenna with the chemical potential of 0 eV and 1 eV, (e) Stokes parameters calculated at 20 nm above the polarizer top surface focusing on the polarizer feed gap \cite{RenRef16}. (f) The near field distributions of the doped graphene monolayer ($\mu_c=1.1$ eV) inserted in between 4 nm $\mathrm{Al_2O_3}$ and 100 nm $\mathrm{SiO_2}$ (as shown in the inset) at the wavelength of 3000 nm, (g) the extinction cross section (ECS) of the metallic dipole antenna with and without the graphene monolayer, (h) the extinction cross section under different chemical potentials \cite{RenRef29}. } \label{RenFig1}
\end{figure*}

Yin et al. proposed the graphene-based two-dimensional (2D) leaky-wave antenna allowing both the frequency tuning and beam steering in the terahertz region. The proposed structure is shown in Fig. \ref{RenFig1}(a), in which the graphene is adopted as the high impedance surface. Through dynamically controlling the gated voltage, the concomitant change of the graphene conductivity not only can effectively alter the reflection phase and resonant frequency over a wide wavelength range but also enable modulation of the radiation pattern of 2D leaky-wave antenna (See Fig \ref{RenFig1}(b)) \cite{RenRef15}.  The \textit{in situ} control of antenna polarization is also desirable in practical applications. Jiang and Sha et al. demonstrated the graphene-based polarizer, in which the graphene monolayer is placed underneath the cross antenna that composed of two perpendicular dipole antennas (See Fig. \ref{RenFig1}(c)) \cite{RenRef16}. The electrostatically tunable chemical potential of the graphene monolayer enables adjustment of the polarization state of the asymmetric cross antenna. The cross antenna at operational wavelength of 6 $\mathrm{\mu m}$ generates a perfect circular state of the reflected beam with the axial ratio of 0.3 dB. Moreover, the axial ratio could be significantly changed by 9 dB when the underneath graphene monolayer is electrostatically biased to achieve the doping level of 1 eV (See Fig. \ref{RenFig1}(d)). The Stokes parameters evaluated within the feed gap indicate that the value of S3 (represents level of circularly polarized state) is dominant compared to that of S1 (represents the differentiate linear polarization) (See Fig. \ref{RenFig1}(e)). This confirms the reflected beam is highly circular in zero doping. The reflected beam is more linearly polarized along the $x$-direction when the chemical potential is tuned to a higher value of 1 eV. Consequently, the reflected wave has been electrically tuned from circular to near-linear states with the axial ratio varying over 8 dB.

The modulations of the radiation pattern, directivity and efficiency of other type antennas such as the Fabry-Perot cavity leaky-wave antenna \cite{RenRef17}, terahertz Yagi-Uda antenna \cite{RenRef18}, the split ring resonator based Fano antenna \cite{RenRef19,RenRef20} and gold nanorod pair antenna\cite{RenRef21} have been demonstrated by shifting the graphene chemical potential under different electrostatic biases. Besides, the graphene has also been adopted as the photoconductive source of radio-frequency antenna and load of terahertz patch antenna. Their time-domain characteristics also show interesting capabilities under dynamic modulations \cite{RenRef22, RenRef23, RenRef24}.

\subsubsection{utilization of graphene plasmon}

The surface plasmon polaritons supported by the graphene monolayer can be dynamically controlled by the electrostatic doping level, which offer a platform to engineer its spectral resonant location and interaction with metal plasmon. The previous reports revealed that the plasmon of graphene and thin metal layer have the comparable performance \cite{RenRef25}. The unidirectional surface plasmon polaritons launcher and the propagation of surface plasmon polaritons have been demonstrated to be electrically controllable through dynamically biasing the source/drain of graphene-based field transistor devices \cite{RenRef26}. Yao et al. numerically investigated the plasmon-induced transparency of the graphene dipole and monopole antenna \cite{RenRef27}. The spectral location and lineshape have been tuned by dynamically controlling the destructive interference between graphene dipole and monopole antenna. Murphy et al. unveiled the anomalously high resonant transmission of graphene-metal antenna and showed the increment of graphene mobility resulting in the enhanced resonant transmission and narrowed bandwidth \cite{RenRef28}. Choy and coworkers have demonstrated the strong mode coupling between the graphene plasmon and metal plasmon in graphene-based dipole antenna. The near-field distribution, resonance frequency, bandwidth, radiation pattern, etc. have been dynamically tuned by the in-phase and out-of-phase couplings between the graphene plasmonic and metallic plasmonic \cite{RenRef29}. The proposed structure adopted a thin metal oxide of $\mathrm{Al_2O_3}$ inserted in between the graphene monolayer and metal dipole antenna for avoiding the quantum tunneling effect between the metal and graphene. The surface plasmon of graphene can be readily seen in the near field profile in Fig. \ref{RenFig1}(f). The results showed that appropriate chemical doping of graphene can induce a strong mode coupling between the graphene plasmon and metal plasmon, which results in a resonance splitting of the graphene-based dipole antenna (See Fig. \ref{RenFig1}(g)). More interestingly, the switch on/off of the mode coupling can be controlled through tuning the graphene chemical potential with electrostatically applied voltage (See Fig. \ref{RenFig1}(h)).  In addition, Halas et. al. proposed the doped graphene through the injection of the plasmon-induced hot electrons, which the prominent carrier density in graphene monolayers, by varying the plasmonic antenna size, incident laser wavelength and laser power density etc., will dynamically shift the Dirac point \cite{RenRevRef1}. In a subsequent work, Fang et. al. demonstrated the plasmonic hot electron generated by the gold nanoparticles can also be tunneled into graphene monolayers, which can be used as the graphene-based photodetector. The tunneling effect of the hot electrons, i.e. the characteristic of the photodetector, can be controlled by incident laser power and bias voltage between the top and bottom electrode \cite{RenRevRef2}.

\subsection{Graphene-based antenna}

\subsubsection{terahertz and microwave antennas}

The terahertz antennas such as the dipole antenna with the size from several to hundreds micrometers fit the small area of experimentally fabricated graphene, which is a suitable platform for extensive study of the graphene-based/modulated antennas \cite{RenRef30, RenRef31, RenRef32, RenRef33}. For instance, the adoption of graphene nanoribbons as the terahertz antenna has been reported to modulate the mode compression factor of surface plasmon polariton and propagation \cite{RenRef34}. Zhang and coworkers presented a dielectric grating with graphene as the leaky-wave antenna. The radiation pattern reconfigurability of the antenna has been realized by controlling the graphene plasmon through the modulation of the applied voltage on the graphene \cite{RenRef35}. Galli et al. proposed the substrate-superstrate configured leaky-wave antenna with graphene as the load to tune the radiation properties \cite{RenRef36}. Recently, Yakovlev and coworkers studied the coverage of the elliptically-shaped graphene monolayer on the two dipole antennas as the surface reactance. The mutual coupling between each dipole antenna can be modulated by the graphene chemical potential resulting in a dynamically controllable radiation pattern \cite{RenRef37}. Chen and coworkers proposed the implementation of a circular antenna to control the beam direction, in which the graphene-metal loop acts as the reflector to dynamically manage the resonant location \cite{RenRef38}. In the another work, the theoretical upper limits on the radiation efficiency and beam steering have been studied for the graphene-based terahertz nonreciprocal antenna \cite{RenRef39}.

\begin{figure*}[!htbp] 
	\centerline{
		\includegraphics[width=6.5in]{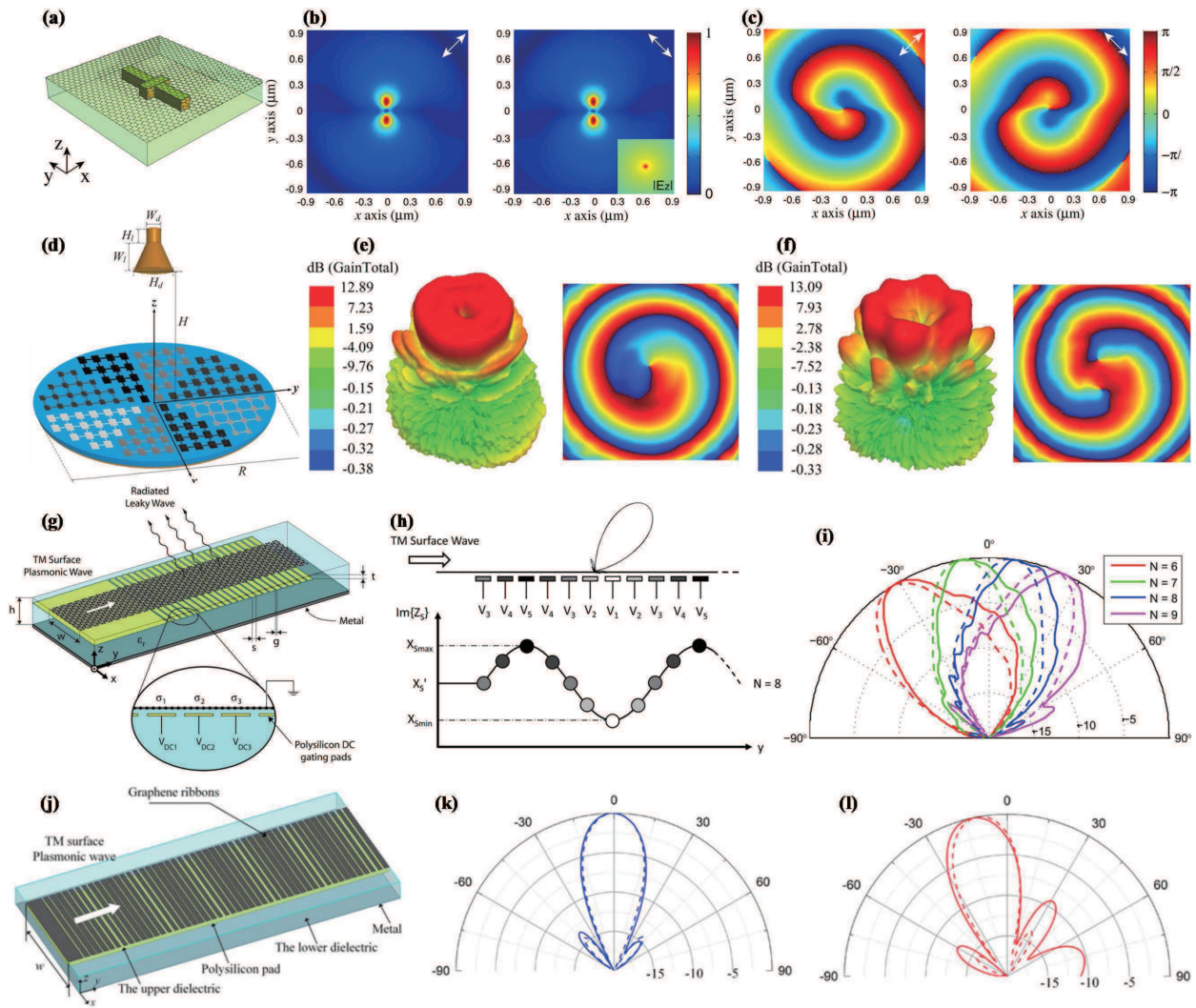}
	}
	\caption{(Color online)  (a) Schematic view of the cross shaped metal antenna, (b) the normalized amplitude and (c) the phase profiles of electric field component $E_z$ excited by the cross shaped antenna under $45^\circ$ and $-45^\circ$ \cite{RenRef40}.  (d) Reconfigurable graphene reflectarray, the dark and bright gray colors correspond to different chemical potentials. The simulated three-dimensional radiation patterns and phase fronts of the vortex radio waves with the modes of (e) $l = -1$ and (f) $l = -2$ \cite{RenRef41}.  (g) Schematic of the sinusoidally-modulated graphene surface reactance operating as a leaky-wave antenna. The polysilicon pads (yellow) are used to modify the graphene conductivity as a function of the applied DC voltage, (h) schematic representation of the relation between the DC bias voltage and graphene reactance, (i) radiation patterns for different values of sampling point $N$ in one modulation period \cite{RenRef48}.  (j) Schematic of the sinusoidally-modulated leaky-wave antenna with nonuniformly gapped graphene nanoribbon, the radiation patterns of the leaky antenna with the chemical potential of (k) 0.5 eV and (l) 0.9 eV \cite{RenRef51}. } \label{RenFig2}
\end{figure*}

Besides the linearly polarized wave, the graphene monolayer also exhibits capabilities to generate vortex waves. Jian and coworkers proposed the cross shaped antenna to generate the plasmon vortex on the coated graphene monolayer (See Fig. \ref{RenFig2}(a)). With the excitation of cross antenna under the linearly polarized light with the directions of $45^\circ$ and $-45^\circ$, the amplitude of electric field component $E_z$ exhibits the sharp change along the rotation direction (Fig. \ref{RenFig2}(b)). The corresponding phase profiles as shown in Fig. \ref{RenFig2}(c) readily revealed features of vortex waves. More importantly, the sign of topological charge of plasmon vortex on graphene can be controlled by the linearly polarized direction of incident light \cite{RenRef40}. Mao and coworkers proposed the reconfigurable graphene reflectarray for the generation of vortex radio wave antenna \cite{RenRef41}. The schematic of the reconfigurable graphene reflectarray is shown in Fig. \ref{RenFig2}(d). The dark and bright gray colors correspond to different chemical potentials. The change of the chemical potential and size of graphene monolayer enable the controllable reflection coefficients and generate the $0$, $\pm1$, and $\pm2$ modes of vortex beams at 1.6 terahertz. The three-dimensional radiation patterns and phase fronts of the vortex beams for the modes of $l = -1$ and $l = -2$ are shown in Figs. \ref{RenFig2}(e) and \ref{RenFig2}(f), respectively. The graphene monolayer also shows potential applications as microwave antennas \cite{RenRef42,RenRef43,RenRef44,RenRef45}. The metal-insulator reversible transitions of graphene via switch on the electrostatic bias for low surface resistance metal and switch off for high surface resistance insulator can efficiently control the microwave antenna's radiation patterns \cite{RenRef46}.

\subsubsection{spatially modulated graphene for antenna applications}

The resonant optical antenna with spatial conductivity pattern has been demonstrated to launch and control the graphene plasmon \cite{RenRef47}. The spatial modulation of the applied bias on graphene has also been studied on various antennas. Perruisseau-Carrier and coworkers proposed the concept of designing a sinusoidally modulated graphene leaky antenna by spatially controlling the gated voltage \cite{RenRef48}. Its configuration is shown in Fig. \ref{RenFig2}(g). The graphene surface reactance can be controlled by sinusoidally modulating the bias voltage applied on the graphene (See Fig \ref{RenFig2}(h)). It results in dynamically tuning the pointing angle and leakage rate of the antenna at a fixed frequency. The radiation pattern with different beam directions can be generated with different sampling points in one modulation period (See Fig \ref{RenFig2}(i)). Alu and coworkers proposed the graphene parallel-plate waveguide loaded with gating pads and demonstrated the spatiotemporal modulation of graphene conductivity. The structure serves as plasmonic isolators and leaky-wave antennas at terahertz frequencies, which allow independent manipulation of graphene properties in both space and time toward the nonreciprocal device \cite{RenRef49}. They further extended the concept to design the infrared beamformer through modulating the graphene with elastic vibrations based on flexural waves. A fast on-off switching of infrared emission and dynamic tuning of radiation pattern, beam angle and frequency of operation have been realized \cite{RenRef50}. Mao and coworkers demonstrated the dynamical tuning of beam scanning of a sinusoidally-modulated graphene leaky-wave antenna with only one biasing voltage \cite{RenRef51}. They have realized the leaky-wave antenna through designing the graphene nanoribbon with the nonuniform gap size, whose configuration is shown in Fig \ref{RenFig2}(j). The nonuniformly gapped graphene ribbons can achieve a sinusoidally-modulated surface reactance to support the leaky-wave operation. With the change of chemical potential from 0.5 eV to 0.9 eV, the radiation beam can be turned from pointing angle of $0^\circ$ to  $15^\circ$. (See Figs. \ref{RenFig2}(k) and \ref{RenFig2}(l))

\subsubsection{graphene-based antenna for photodetection}

In the terahertz and visible light regions, the rectifying antenna is used to convert electromagnetic energy into direct current electricity that has promising applications in photodetector and wireless communication systems \cite{RenRef52,RenRef53,RenRef54,RenRef55}. The introduction of graphene for modulating the responsivity of photodetector and rectenna also have received intensive attentions. Capasso and coworkers have demonstrated a metallic antenna with simultaneously improved light absorption and photocarrier collection in graphene-based detectors. The metallic antenna serves as the light trapping structure for substantial absorption enhancement and the electrode for efficient carrier collection. The resultant responsivity of the metallic antenna-assisted graphene detector has been enhanced over 200 times \cite{RenRef56}. Halas and coworkers also proposed a graphene antenna as the photodetector, in which the antenna is sandwiched between two graphene monolayers. The plasmon of the antenna will benefit the direct excitation of graphene electrons and also induce hot electrons that are subsequently transferred to graphene. They cooperatively enhance the photocurrent with 8 times \cite{RenRef57}. For the wireless communication, various graphene-based patch antennas with low profiles and small sizes have also been comprehensively studied. The adopted graphene monolayer offers the possibilities to actively control the directivity, and bandwidth of the patch antenna \cite{RenRef58,RenRef59,RenRef60,RenRef61,RenRef62,RenRef63}.  The graphene monolayer has also been reported to fabricate the radio frequency identification tag \cite{RenRef64,RenRef65,RenRef66}, and realize the multiple-input multiple-output antenna system with tunable beamwidth and radiation direction \cite{RenRef67,RenRef68}.

\section{Graphene plasmonic modulators at terahertz and infrared}
Modulating the light intensity, phase, and polarization plays a critical role in most optical imaging and communication systems. Compared to modulators operating at telecommunication wavelengths that are predominantly used for fiber communications, terahertz or infrared spatial light modulators are uniquely targeted to environment/health monitoring, navigation, and surveillance through scattering foggy or dusty environments.

Due to its large conductivity swing, graphene has attracted lots of attention as a material for terahertz modulators \cite{BRef114, HeRef115, TredRef116, TamagnoneRef117}. Since 2010s, there have been many reports for a large modulation depth and broadband operation in the terahertz regime by actively controlling the carrier concentration of a graphene layer and via the integration of graphene with metamaterials \cite{Otsuji,BRef114, HeRef115, TredRef116, TamagnoneRef117, ChenRef118, RahmRef119, ChenRef120, TanotoRef121, LeeRef122, ZhouRef123, AndryieuskiRef124, AminRef125, VasicRef126, YangRef127, BludovRef128, LiuRef129}. In 2012, Sensale-Rodriguez et al. \cite{Sensale-RodriguezRef130} developed a broadband graphene terahertz electro-absorption modulator. By applying a voltage between a graphene layer and a silicon substrate, in a large-area field-effect transistor-like configuration, the Fermi level in the graphene film can be well controlled. Although the modulation depth achieved using this technique is remarkable, owing to the one-atom thickness of graphene as the active material, the overall device performance is still far away from application expectations. However, there are several ways to enlarge the modulation depth. One of them is to utilize the reflection mode structures \cite{SensaleRef132}. Integrating graphene into such structures can boost the absorption in graphene due to electric field enhancement when the distance between graphene and a reflector (e.g. a metallic layer coated on the back of the substrate to which graphene is transferred) is matched to an odd multiple of a quarter wavelength \cite{SensaleRef132}. The experimental reflectance as a function of bias voltage shows that at the frequency at which the substrate thickness is matched to an odd multiple of a quarter wavelength, i.e. $0.62$ terahertz maximum reflectance modulation is attained, and $>64\%$ amplitude modulation with $2$ dB insertion loss is possible in this structure. However, owing to the characteristics of its geometry, the response of this structure is narrow band.

\begin{figure*}[!htbp] 
	\centerline{
		\includegraphics[width=6.5in]{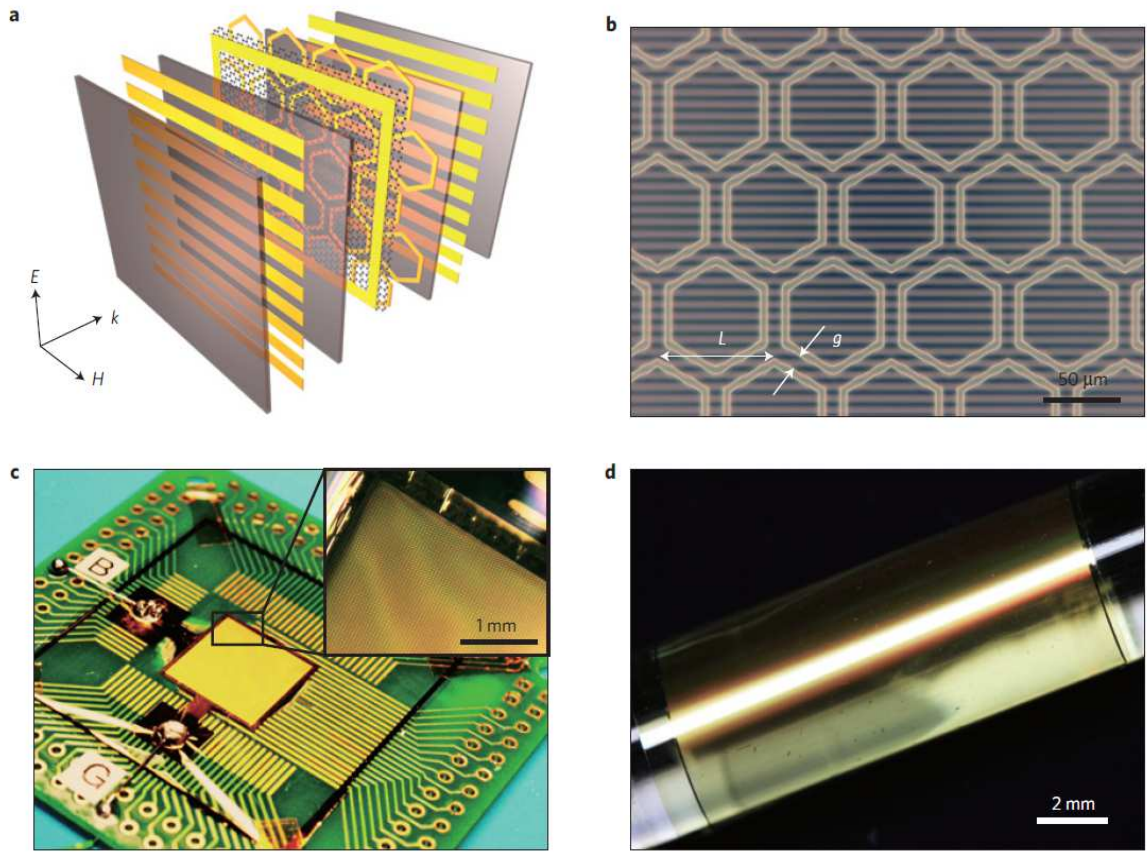}
	}
	\caption{(Color online) Schematic view and device images of gate-controlled active graphene metamaterials. (a) Schematic rendering of a gate-controlled active graphene metamaterial composed of a SLG deposited on a layer of hexagonal metallic meta-atoms (a unit cell of $ L=60\,\,\mathrm{\mu m}$ and $g=5\,\,\mathrm{\mu m}$) and top/bottom EOT electrodes embedded in a dielectric material of $4$ thickness. Metallic patterns and dielectric materials are represented by yellow and grey, respectively. The polarization of the incident terahertz wave is perpendicular to the line electrode, as indicated by the arrows.  (b) Optical micrograph of the fabricated gate-controlled active graphene metamaterial without the top electrode.  (c) Fully integrated gate-controlled active graphene metamaterial attached to a drilled PCB for THz-TDS measurement (B, connected to bottom EOT electrode; G, connected to graphene layer). Inset: magnified view of the gate-controlled active graphene metamaterial.  (d) Optical image of the fabricated large-area metamaterial wound round a glass rod, showing its high degree of flexibility. Source: Reprinted with permission from Ref. \cite{LeeRef147}. Copyright from 2012 Nature Publishing Group.} \label{ShanFig5}
\end{figure*}

On the other hand, terahertz modulation is also attainable in graphene plasmonic structures. In this regard, Ju et al. \cite{JuRef134} proposed a new device design composed of stripes of graphene. By changing the carrier density in graphene, it is possible to shift the geometric plasmonic resonance of the device. One step further is to integrate graphene with metamaterials. It takes advantage of the electric field enhancement due to the metamaterial in order to boost the light-matter interaction in graphene \cite{ZhangRef137, DingRef138, WangRef139, GrzekiewiczRef140, WangRef141, BornRef142, WangRef143, AnsellRef144,MaoRef145, HeRef146, LeeRef147, YanRef148}. In 2012, Lee et al. \cite{LeeRef147} reported a graphene metamaterial modulators with a very large modulation depth  ($86\%$) , but at the expenses of a large insertion loss ($>10$ dB).  As depicted schematically in Fig. \ref{ShanFig5}, two extraordinary transmission electrodes are placed on both sides of this layered structure so as to gate the active terahertz graphene metamaterial (Fig. \ref{ShanFig5}(a)). And the meta-atoms composed of a hexagonal metallic frame or asymmetric double split rings (aDSRs) exhibiting Fano-like resonance are periodically arranged (Fig. \ref{ShanFig5}(b)) \cite{LeeRef147}. An optical micrograph of the fabricated graphene metamaterial attached to a printed circuit board (PCB) is shown in Fig. \ref{ShanFig5}(c). As shown in Fig. \ref{ShanFig5}(d), the fabricated active graphene metamaterial is large-area ($15\times15\,\,\mathrm{mm}^2$), fully flexible and free-standing without the thick base substrate that is generally required for semiconductor-based terahertz modulators \cite{LeeRef147}.

\begin{figure*}[!htbp] 
	\centerline{
		\includegraphics[width=6.5in]{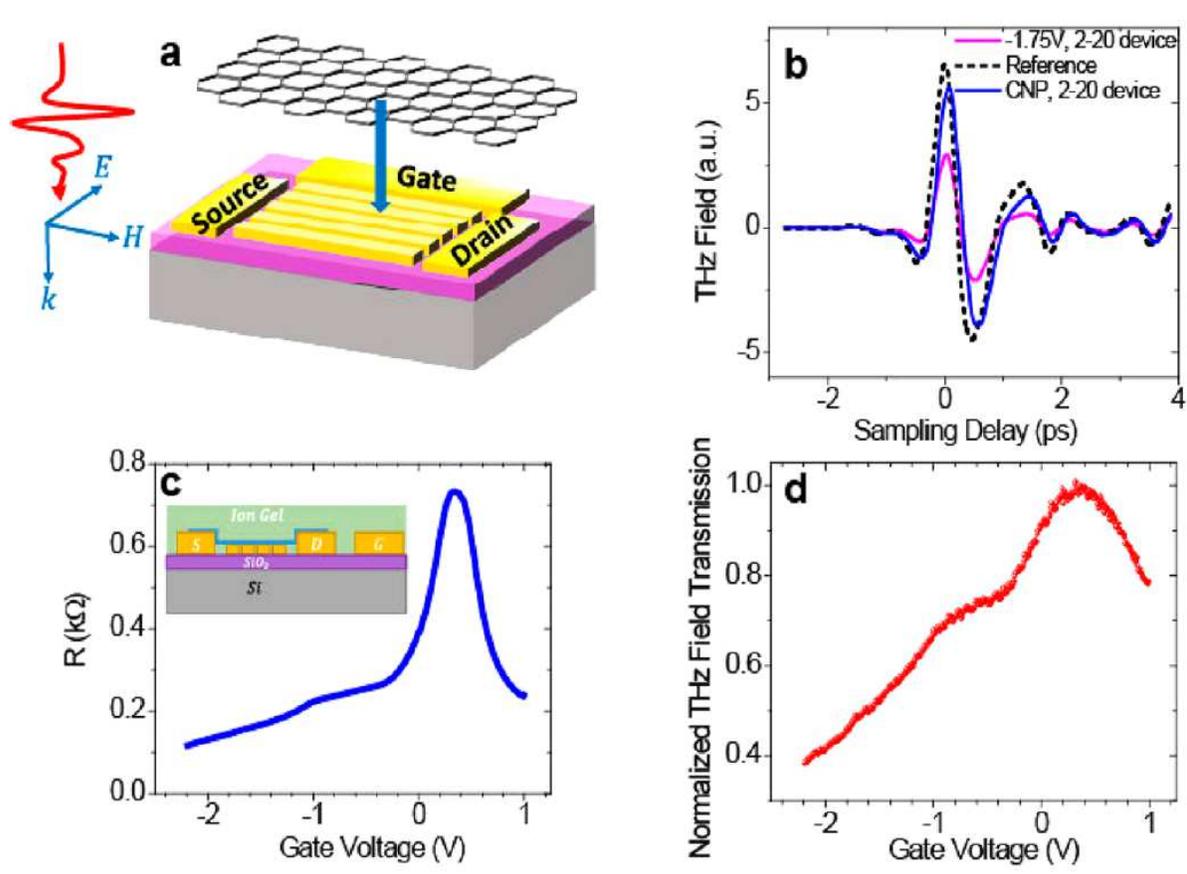}
	}
	\caption{(Color online) Terahertz modulation by the graphene/metasurface hybrid device. (a) Schematic representation of the experiment setup for measurement as well as the hybrid device configuration. We fabricate the hybrid device by transferring a single layer graphene on top of a gold slit device of width $2\,\,\mathrm{\mu m}$ and period $20\,\,\mathrm{\mu m}$. The incoming terahertz wave is polarized perpendicular to the slit orientation. (b) Transmitted terahertz waveforms for a reference sample (dashed black trace), the hybrid device with graphene at charge neutrality point (CNP, solid blue trace), and the hybrid device with graphene at gate voltage -1.75 V (solid magenta trace). (c) Resistance measured for this hybrid device shows charge neutrality point at the gate voltage of 0.33 V. The side view of a schematic representation of our ion-gel gated device is shown in the inset, where the blue line represents graphene. (d) Simultaneously measured terahertz field transmission of this hybrid device (normalized to the transmission when graphene is at charge neutrality point). Source: Reprinted with permission from Ref. \cite{ShiRef151}. Copyright from 2015 American Chemistry Society.} \label{ShanFig6}
\end{figure*}

From the experimental results of the phase shift versus applied bias, a $65^{\circ}$ phase shift is achieved at a frequency of $0.9$ THz. Here we would like to note that in these structures there is always a tradeoff between the modulation depth, insertion loss and the bandwidth of modulation. Later in 2014, a low-bias terahertz modulator based on the integration of split-ring resonator and a graphene layer was further proposed by Degl'Innocenti \cite{DeglRef149}. The modulation in a broad spectral range ($2.2-3.1$ terahertz) with a modulation depth of up to $18\%$ was achieved. In 2012, a  metamaterial terahertz modulator based on periodic gold slit arrays with graphene as an active load of tunable conductivity was proposed in \cite{NovitskyRef150} and further experimentally demonstrated in 2015 by Shi et al. in \cite{ShiRef151}. A sketch of this structure is schematically depicted in Fig. \ref{ShanFig6}(a). The field enhancement in the graphene layer is attained via the coherent radiation of the enhanced in-plane local-field in the slits, thus preserving the broadband terahertz modulation of bare graphene in the hybrid device. Thus this structure is capable of efficiently coupling terahertz radiation while exhibiting tunability of the terahertz transmission over a broad frequency range as shown in Fig. \ref{ShanFig6}(b) and Fig. \ref{ShanFig6}(d). The schematic side view of the ion-gel gated device is shown as the inset of Fig. \ref{ShanFig6}(c) \cite{ShiRef151}. Under this scheme strong terahertz absorption and a large local field enhancement exist, leading to a modulation depth of $90\%$. It is also possible to improve the modulation depth in these structures by increasing the number of graphene layers. It should be noted that as the number of layers increases, the insertion loss also increases. Furthermore, in 2015, Wu et al. \cite{WuRef152} reported a broadband terahertz modulator based on graphene/ionic liquid/graphene sandwich structures, in which the obtained modulation depth reached up to $99\%$ in the frequency range $0.1-2.5$ terahertz.

One interesting approach towards high-performance graphene modulators is via the utilization of hybrid systems. By optically pumping the structure and properly engineering the device, it is possible to obtain significant modulation depth \cite{WeisRef153,WenRef154}. A graphene terahertz modulator based on an optical pumping approach was demonstrated by Wen et al. in \cite{WenRef154}, which was based on single-layer graphene on germanium. The modulator was driven optically with a low power $1.55$ $\mathrm{\mu m}$ laser, in which the main mechanism to enable the terahertz modulation is the third-order nonlinear effect \cite{MikhailovRef155,YangRef156} in the optical conductivity of the graphene. The modulator shows wide-band modulation in the range $0.25-1$ terahertz with a modulation depth up to $94\%$. In 2015, Liang et al. \cite{LiangRef158} demonstrated that $>94\%$ modulation depth can be attained via the monolithic integration of a surface-emitting terahertz quantum cascade laser featuring a concentric circular grating with modulation speeds up to $110$ MHz. The improvement in performance and speed is achieved as a result of the small area of the device and the stronger interaction of the terahertz radiation with the graphene layer, due to monolithic integration of modulator and source. Moreover, a graphene-silicon hybrid device was experimentally demonstrated by employing optical excitation and direct-current bias together as the actuation mechanism \cite{LiRef159}. This new graphene-silicon configuration device with a 'diode' behavior is actuated under simultaneous continuous wave photoexcitation and direct-current bias voltage, which exhibited a relatively large modulation depth up to $83\%$ at a low gate bias voltage of $4$ V.

\section{Graphene detectors at terahertz}
Although several room temperature field effect transistors have been shown capable of detecting terahertz waves with very fast response and high responsivity in the $0.3-3$ terahertz range based on InAs nanowire field-effect transistors \cite{TredRef160,SizovRef161,SHANRef162}, graphene plasmonics as one exciting frontier offers lies in harvesting low energy photons for robust photodetection at terahertz at room temperature, which can potentially overcome the limitations of conventional semiconductor terahertz devices. It oughts to be noted that, since the thermal diffusion of hot carriers instead of lattice heating is responsible for generating the current, photo-thermoelectric graphene detectors can achieve fast frequency response. It could be generalizable to terahertz frequencies, where carriers in graphene have an even stronger response \cite{SHANRef162}. Conventional terahertz and sub-terahertz detection systems based on incoherent (pyroelectric, Golay cell, Si bolometers) or coherent (heterodyne mixers) approaches are either very slow ($~100$ Hz modulation frequency) or require deep cryogenic cooling, while those exploiting fast nonlinear electronics (Schottky diodes) show a significant drop of performance above $1$ terahertz  \cite{SizovRef161}.Traditional detectors based on interband transitions (e.g., mercury cadmium telluride) require cryogenic cooling to suppress dark-current noise, especially when operating at long wavelength infrared wavelengths. Moreover, device miniaturization and their heterogeneous integration with silicon or germanium platforms remain challenging. Uncooled thermal detectors such as thermopiles or bolometers usually possess large device footprints, low responsivities and slow speeds due to thermal time constants being in the millisecond range. These drawbacks have significantly impeded their applications in thermal imaging and free space communications.

Recently, it has been generally recognized that graphene can pave the way to robust and cheap terahertz detectors operating at room temperature based on the Dyakonov and Shur scheme \cite{TredRef160,VicarelliRef167}. Graphene has a very high carrier mobility, even at room temperature \cite{Neto}. Furthermore, it supports plasma waves that are weakly damped in high-quality samples \cite{Grigorenko,Jablan}. Thus, single-layer and bilayer graphene field-effect transistor plasma-based photodetectors could outperform other terahertz detector technologies. In the last couple of years, various graphene based structures have been proposed i.e. single-layer or bilayer graphene-insulator-graphene heterostructure tunnel field-effect transistor \cite{FeenstraRef168,RyzhiiRef169,MishchenkoRef170,FallahazadRef171}, lateral single-layer or bilayer graphene field-effect transistor \cite{TomadinRef173}, vertical cascade multiple graphene layer structures \cite{RyzhiiRef174}, etc., for resonant detections. Moreover, some interesting works demonstrate that such graphene-insulator-graphene heterostructure tunnel field-effect transistor structures can show strong resonant terahertz detections, compared to ungated tunnel field-effect transistor structures, and interesting readers can refer to Refs. \cite{SHANRef162,TredRef160,VicarelliRef167,SensaleRef175}.

Graphene has also been proposed for the broadband terahertz detection by employing a simple top-gate antenna-coupled configuration for the excitation of overdamped plasma waves in the channel of a graphene field-effect transistor [65]. Several further studies on similar split-bow tie antenna have also been reported to improve the coupling and thereby responsivity at $0.34$ terahertz and 0.6 terahertz \cite{ZakRef176,ShanRef177,Knap2013}.
On the other hand, some interesting results were also obtained for detections at $1.63$ and $3.11$ terahertz using graphene field-effect transistors, whose drain and source contact leads worked as antennas for the incoming terahertz radiation \cite{SongRef179}.  Moreover, several photodection devices have also been demonstrated in the mid-infrared and far infrared regimes \cite{Freutag180,YanRef181}. In Ref. \cite{YanRef181}, the bilayer graphene was used as the bilayer graphene small electron heat capacity; and weak electron-phonon coupling/interaction creates a bottleneck in the heat path, thus leading to a temperature dependent resistance induced by the large change in electron temperature, which can be then converted to a detectable electrical signal. The developed detector exhibits a voltage responsivity about $2\times10^5\,\,\mathrm{V}\,\mathrm{W}^{-1}$ and an electrical noise-equivalent power about $33$ $\mathrm{fW\,Hz}^{-1/2}$ at 5 K with a very high intrinsic speed of the device ($>1$ GHz at $10$ K). Besides, room temperature terahertz detection using the photothermoelectric effect in graphene has also been shown experimentally \cite{CaiRef182,MuravievRef183} .

Besides, the graphene monolayers have also been widely explored as other functional devices such as the transistors, sensors, memorizers, etc. The graphene-based transistors have shown great promising to reach a smaller channel length and higher speed, which offer the opportunities to outperform the existing devices and become an option for post-silicon electronics \cite{RenRevRef4,RenRevRef5,RenRevRef6}. Recently, Sahakian et. al. demonstrated the cascaded spintronic logic with graphene nanoribbon based transistors. The proposed computing systems permit terahertz operation speed and two orders of magnitude decrement in power-delay product as compared to the current cutting-edge microprocessors \cite{RenRevRef7}. The unique characteristics of the graphene monolayers are also of critical use in the next-generation memorizers due to its superiorities in bit density, energy efficiency and long data retention time \cite{RenRevRef8}.  The graphene monolayers have been applied in the resistive random memory, phase-change memory, spin-transfer-torque magnetic random access memory and ferroelectric random access memory etc., in which the graphene can be used as the memory electrode due to its flexibility and transparency, and interfacial engineering layer due to the block of atomic diffusion, reduction of power consumption etc. \cite{RenRevRef9,RenRevRef10,RenRevRef11,RenRevRef12,RenRevRef13}.  The graphene-based non-volatile memory devices has been systematically summarized in previous works \cite{RenRevRef8,RenRevRef14}. 
In addition, the graphene field effect transistor can also be used as the smart sensor, in which the advantages of graphene monolayers such as the large surface-to-volume ratio, uniquely tunable optical properties, excellent carrier mobility, and exceptional electrical and thermal properties, are suitable for sensing \cite{RenRevRef6,RenRevRef15}.  There are several works on the graphene-based strain sensor, electrochemical sensor, biosensor and electrical sensor (including the temperature sensing, photodetector, etc.), in which the graphene reveal great potential to achieve the low detection limit and high sensitivity \cite{RenRevRef15,RenRevRef16,RenRevRef17,RenRevRef18}. Notably, the graphene-based sensors have the low influence on the surrounding environment as compared to the metal-based sensors. Nalwa et. al. have detailed reviewed the flexible graphene-based wearable gas and chemical sensors in the detection of various hazard gases, toxic heavy ions, and volatile organic compounds etc.  \cite{RenRevRef19}.  In addition, the decorations of the graphene monolayers with metal nanostructures permit the large enhancement of the chemical detection capabilities \cite{RenRevRef20}, the intensified surface-enhanced Raman scattering has been reported with metal nanoparticle/graphene/metal film system  \cite{RenRevRef21,RenRevRef22,RenRevRef23}. Consequently, the graphene monolayers have already shown great potential in functional devices and promised in future practical applications.

\section{Conclusion}
Graphene represents an emerging frontier that brings together multidisciplines of physics, material science, electronics, photonics and terahertz plasmonics. It has uncovered exciting prospects in these fields, where novel technologies and solutions are still being developed.
From the device and system viewpoint, graphene is a two-dimensional layered material with an intrinsically passivated surface. It is thus amenable to monolithic integration with conventional silicon-based benchmarks as well as other important functional materials, which makes graphene remarkably advantageous over bulk materials whose heterogeneous integration with other materials usually induces defects and deep level trapping centers at the interface.

Presently graphene has been suffering some demanding challenges that must be overcome before making an impact on industry-standard device applications. Among many others, the high quality synthesis of large-scale graphene is of paramount importance and will have the most positive impact on this field. Specifically, the improvements in graphene's crystal quality (i.e., carrier mobility) can significantly prolong the lifetime of graphene plasmons, which is strongly correlated with various device performance metrics, such as the light-matter interaction strength, absorbance, sensitivity,phase shifts, spectral resolution, modulation efficiency, detectivity, and so forth.

Moreover, there remains a most important requirement to gain a detailed and fundamental physical understanding of the hot carrier dynamics and transports in graphene since they are indispensable for efficiently developing high-performance graphene-based informatic functional devices. Furthermore, the nonlocal, nonlinear and quantum effects existing in deep subwavelength graphene nanostructures deserve to be intensively investigated in future as well, which will open up a new platform for quantum information, communication, and measurement. The light-matter interaction will occur at both the atomic/molecular level and single-photon level. Additionally, the graphene based multi-functional on-chip devices operating at microwave, terahertz, and visible light frequencies are expected to be invented.

In summary, the various physical and chemical properties of graphene make it well suited for integrated, multifunctional and compact devices and systems, yet its potential has not been fully reached. In this review, we offered a brief on graphene based transmission lines, antennas, modulators, and detectors covering from microwave frequency band to terahertz or infrared band. We believe that future progresses on graphene-based functional devices and microsystems will lead to proliferation in both fundamental investigations and applied technological products for imaging, communication, environmental monitoring, medical diagnostics, spectroscopic studies etc.

\begin{acknowledgments}
This research work is supported by the National Key R\&D Program of China (Grant No. 2016YFE0204200), the Natural Science Foundation of China (61701003) and Natural Science Research Foundation of Anhui Province (No. 1808085QF179). This work is also supported by the Research Grants Council of Hong Kong GRF 17210815, NSFC 61271158, HKU Seed Fund 201711159228, AOARD FA2386-17-1-0010, and Hong Kong UGC AoE/P-04/08.

\end{acknowledgments}

\end{document}